\documentclass[prd, aps, nofootinbib,preprintnumbers, showpacs,superscriptaddress,twocolumn]{revtex4}

\usepackage{latexsym}
\usepackage{amssymb}
\usepackage{amsfonts}
\usepackage{amsmath}
\usepackage{bm}
\usepackage[dvips]{graphicx}
\usepackage{color}
\usepackage{times}

\newcommand{\sech}{\operatorname{sech}}

\allowdisplaybreaks

\begin{document}

\title{3+2 Cosmology: unifying FRW metrics in the bulk}

\date{\today}
\label{firstpage}

\author{Carles Bona}
\affiliation{
   Universitat Illes Balears, Institute for Computational Applications with Community Code (IAC3), Palma, Spain
}
\affiliation{
 Institute for Space Studies of Catalonia, Barcelona, Spain
}

\author{Miguel Bezares}
\affiliation{
Universitat Illes Balears, Institute for Computational Applications with Community Code (IAC3), Palma, Spain
}

\author{Bartolom\'{e} Pons-Rullan}
\affiliation{
  Universitat Illes Balears, Institute for Computational Applications with Community Code (IAC3), Palma, Spain
}

\author{Daniele Vigan\`{o}}
\affiliation{
 Universitat Illes Balears, Institute for Computational Applications with Community Code (IAC3), Palma, Spain
}

\begin{abstract}
The Cosmological Problem is considered in a five-dimensional (bulk) manifold with two time coordinates, obeying vacuum Einstein field equations. The evolution formalism is used there, in order to get a simple form of the resulting constraints. In the spatially flat case, this approach allows to find out the general solution, which happens to consist in a single metric. All the embedded Friedmann-Robertson-Walker (FRW) metrics can be obtained from this 'mother' metric ('M-metric') in the bulk, by projecting onto different four-dimensional hypersurfaces (branes). Having a time plane in the bulk allows to devise the specific curve which will be kept as the physical time coordinate in the brane. This method is applied for identifying FRW regular solutions, evolving from the infinite past (no Big Bang), even with an asymptotic initial state with non-zero radius (emergent universes).
Explicit counter-examples are provided, showing that not every spatially-flat FRW metric can actually be embedded in a 3+2 bulk manifold. This implies that the extension of the Campbell theorem to the General Relativity case works only in its weaker form in this case, requiring as an extra assumption that the constraint equations hold at least in a single 4D hypersurface.
\end{abstract}

\maketitle


\section{Introduction}

Since the pioneering work of Kaluza-Klein, adding extra dimensions to the standard (4D) spacetime has widely been adopted as an strategy in the quest for unification. In modern particle physics, the resulting higher-dimensional 'bulk' spacetime is considered to be the basic physics scenario, in which gravity acts, although ordinary matter and fields are supposed to be confined in 4-dimensional 'branes'~\cite{Randall-Sundrum, Maartens, Maartens_living}. This idea has also been implemented in Cosmology, after the pioneering work of Ponce de Leon~\cite{JPL88} (see~\cite{Langlois,Brax} for a review).
Another example is provided by the well known superstring theory result: the 10+1 M-theory encompasses the previously known 9+1 theories, which can be interpreted as just different projections of the same (unifying) theory~\cite{Witten}.

The Kaluza-Klein work, however, unified not only forces (gravity and electromagnetism), but also matter and geometry. Matter fields in the standard 4D spacetime arise from metric coefficients in a vacuum 5D bulk manifold. This spacetime-matter (STM) unifying approach has shown to be fruitful (see~\cite{Overduin} for a review), and it is also being applied to Cosmology~\cite{JPL, Camera, Madriz, Bellini, Georgalas}. In some of these papers different projections of the same bulk metric are considered~\cite{JPL, Camera}, so that different FRW metrics are obtained just by projecting onto different branes.

In this paper, we will adopt a top-down approach: we will consider a five-dimensional bulk manifold with three space coordinates plus two time coordinates. We choose the 3+2 signature, instead of the 4+1 alternative, because the explicit embedding formulae for any FRW metric into 5D Minkowski space are already known~\cite{Rosen}. Moreover, one can wonder why only half of the solutions in the pioneering work of Ponce de Leon~\cite{JPL88} where embedded in a 3+2 bulk, whereas all the eight solutions where embedded in the 4+1 alternative bulk. We interpret this as a clear hint that the extension to the pseudo-Riemannian case of the classical Campbell theorem~\cite{Campbell, Seahra} deserves a closer look in the 3+2 context, where the extra dimension is time-like.

The 3+2 approach implies dealing with two time coordinates. This idea has yet been used in this context~\cite{Bars, Li, Wesson} (see~\cite{Bars book} for a review). The presence of two time coordinates raised some causality concerns, after the G\"{o}del claims about closed time-like curves and the possibility of time travel~\cite{Godel}. Although some authors extended these concerns to more general spacetimes~\cite{Tipler, Got, Visser}, it has been shown that the possibility of closed timelike curves is not inherent to the two-times scenario, being rather due to the point identification involved in the compactification process~\cite{Cooperstock}. Our approach is that matter fields appear only on the projected branes, where there is just one time coordinate left. In this way will not require any compactification mechanism, so we are on safe ground.

There is also a bonus for using two time coordinates: we get a time plane instead of a time line. As we will see, this provides extra symmetries, arising from the group of conformal transformations in the plane, that can be used for solving the vacuum field equations in the bulk. Moreover, we get additional freedom in the choice of the time coordinate which will be actually projected onto a given brane: it can be selected by just drawing a suitable line in the time plane. This flexibility has been crucial in order to obtain our results.

The paper is organized as follows: in section~2, we express the cosmological problem in the language of the evolution formalism. This is the 4+1 extension of the well known 3+1 General Relativity formalism (see for instance~\cite{4+1}). We take advantage of the symmetry properties in the time plane in order to reduce the full set of vacuum Einstein equations to only two equations, for a generic form of the space-homogeneous metric with two functions of the time coordinates. In section~3, we focus on the spatially flat ($k=0$) case, in which the above mentioned symmetries can be fully exploited. Making use of the remaining coordinate freedom, we obtain the general solution, which consists (except for the trivial flat case) in a single metric, which we will call 'M-metric'. Every embedded ($k=0$) FRW cosmological models can be obtained by projecting this M-metric in a suitable four-dimensional brane. This is the multiple projection feature already detected in~\cite{JPL}: the novelty here is that a single 'mother' metric can be taken as the starting point. In section~4 we show how this multiple projection capability can be specially fruitful in a bulk with two time coordinates. Just drawing suitable lines in the time plane allows one to design universe models with some specific properties. We will focus in regular models (with no big bang), by providing many simple examples with diverse properties. The infinite past limit can be either a big bang singularity (just as a limit, never reaching it) or a finite radius universe (emergent models). All cases start with an inflationary phase, followed by a deceleration phase. Some cases keep expanding without bound, whereas others approach asymptotically some stationary state.
Finally, in section~5 we take a closer look to the modern extensions of the Campbell theorem. We will provide explicit counter-examples to the common belief that any 4D metric can be embedded in a 5D manifold, with (4+1) or (3+2) signature. Our results will show that the Campbell theorem does not ensure the embedding of every 4D metric into a Ricci-flat 5D manifold when the extra dimension is time-like (3+2 signature).


\section{5D Cosmological framework: evolution formalism}

We will consider here 5-dimensional (5D) vacuum metrics, where the extra time coordinate is labeled by $\psi$. In our case, where we assume space homogeneity and isotropy, this 'bulk' metric would read
\begin{equation}
    ds^2 = - \alpha^2(\psi,t)\, d\psi^2 - N^2(\psi,t)\, dt^2 + R^2(\psi, t)\;\gamma_{ij}\; dx^i\,dx^j\,,
\end{equation}
where the three-dimensional metric $\gamma_{ij}$ is of constant curvature, that is
\begin{equation}
    ^{(3)}\!R_{ij} = 2k\;\gamma_{ij}\;\;\;\;\;k\,=\,0,\pm 1\,.
\end{equation}

Let us look now at the time plane. We can take advantage from the fact that any Riemannian
2D metric is conformally flat in order to simplify the bulk metric form, namely
\begin{equation}\label{bulk metric}
    ds^2 = - A^2(\psi,t) (\,d\psi^2 + dt^2\,) + R^2(\psi, t)\;\gamma_{ij}\; dx^i\,dx^j\,.
\end{equation}
In this way the symmetry in the $(\psi,t)$ plane (time plane) is manifest.
On every constant $\psi$ hypersurface we will of course recover a FRW line element, namely
\begin{equation}\label{FRW}
    - A^2(\psi,t) \,dt^2 + R^2(\psi, t)\;\gamma_{ij}\; dx^i\,dx^j \equiv g_{ab}\; dx^a\,dx^b\,,
\end{equation}
where $a,b\,=\, 1,2,3,4$.

We can now consider the 4+1 decomposition of the vacuum Einstein equations for the bulk metric (\ref{bulk metric}). It is usually cast as a system of ten evolution equations (along the $\psi$
lines) for the extrinsic curvature $K_{ab}$ of the projected metric (\ref{FRW}), namely
\begin{equation}\label{extrinsic curvature}
    \partial_\psi ~ g_{ab}  \equiv - {2\,A}\,K_{ab}\,,
\end{equation}
plus five constraints (not involving $\psi$ derivatives) of the basic fields $(\gamma_{ab},K_{ab})$.
These numbers, however, are actually reduced by the spatial isotropy assumption. To begin with, the extrinsic curvature can be explicitly computed:
\begin{equation}\label{extrinsic curvature result}
    K_{ab}\,=\, - \frac{R'}{AR}\, (g_{ab}+u_a\,u_b) + \frac{A'}{A^2}\, u_a\,u_b\,,
\end{equation}
where the primes stand for $\psi$ derivatives and $u^a$ is the future-pointing time unit vector (the FRW metric four-velocity)
\begin{equation}\label{fourvelocity}
    u^a\,=\, \frac{1}{A}\; \delta^a_{(t)}\,,
\end{equation}
which of course verifies
\begin{equation}\label{uderivatives}
    \nabla_a\,u_b\,=\frac{\dot{R}}{AR}\, (g_{ab} + u_a u_b)\,,
\end{equation}
where the dots stand for $t$ derivatives and $\nabla$ is the covariant derivative operator in the projected hypersurface.
This means that the symmetric tensor $K_{ab}$ has only two (instead of ten) independent components, so that the evolution equations
\begin{equation}\label{evolution eqs.}
    \partial_\psi~{K_a}^b = -\nabla_a\,\partial{\,^b}\,A
    + A\;   [\,^{(4)}\!{R_a}^b + trK \,{K_a}^b\,]\,,
\end{equation}
contain just two independent conditions as well.

A similar reduction occurs in the vector constraint, namely
\begin{equation}\label{vector constraint}
    \nabla_b\,[\,{K_a}^b - trK \,{\delta_a}^b\,] = 0\,,
\end{equation}
has four components, but the space direction contribution vanishes identically due to the spatial isotropy. The only non-trivial contribution can be written, allowing from (\ref{extrinsic curvature result}), in a very simple form:
\begin{equation}\label{vector constraint reduced}
    A\,\dot{R}'\,=\, A'\dot{R} + \dot{A}R'\,,
\end{equation}
where the dots stand for time derivatives. The remaining (scalar) constraint equation reads
\begin{equation}\label{scalar constraint}
    {K_a}^b{K_b}^a - (trK)^2 =\, ^{(4)}\!R\,,
\end{equation}
where $^{(4)}R$ is the scalar curvature of the FRW metric, that is
\begin{equation}\label{4D scalar curvature}
    ^{(4)}R = \rho-3p = \frac{6}{R^2}\,[\,k + \frac{1}{A}\,\partial_t(\frac{R\dot{R}}{A})\,]\,.
\end{equation}

At this point, we must note that the 5D Bianchi identities ensure that the constraint equations (\ref{vector constraint}, \ref{scalar constraint}) are first integrals of the evolution equations (\ref{evolution eqs.}). In our case, as there are just two independent evolution equations for two independent constraints, we can conclude that the set of two conditions (\ref{vector constraint reduced}, \ref{scalar constraint}) amounts to the full set of vacuum field equations for the bulk line element (\ref{bulk metric}). In spite of that, the evolution equations (\ref{evolution eqs.}) can still be of some use. The space components contribution, for instance, can be expressed in a quite simple form
\begin{equation}\label{evolution eqs reduced}
    (R^3)''\,+ (R^3)\ddot{~} = -6k\,A^2R\,,
\end{equation}
which can give a clue in order to solve the basic system (\ref{vector constraint reduced}, \ref{scalar constraint}).


\section{General solution for the $k=0$ case: The M-metric}

Let us note that the form (\ref{bulk metric}) does not exhaust coordinate freedom, as any conformal transformation in the time plane will preserve this form of the line element. Allowing for the fact that the conformal group in the plane has infinite dimension, it follows that the metric coefficient $A(\psi,t)$ is strongly coordinate-dependent. To be more specific, we can consider any analytic function $\lambda(\psi,t)$ in the time plane in order to get
\begin{equation}\label{conformal transf}
    A^2(\psi,t) (\,d\psi^2 + dt^2\,) = \tilde{A}^2(\lambda,\phi) (\,d\phi^2 + d\lambda^2\,)\,,
\end{equation}
where $\phi(\psi,t)$ is the harmonic conjugate of $\lambda$.

Allowing for (\ref{evolution eqs reduced}), it follows that, in the spatially flat case ($k=0$), the function $u \equiv R^3$ is harmonic. This means that we can take $u(\psi,t)$, and its harmonic conjugate $v(\psi,t)$ as the time plane coordinates, that is taking~\footnote{We are assuming here that u is not constant. If it is, then equations  (\ref{vector constraint reduced}, \ref{scalar constraint}) are identically satisfied. $A(\psi,t)$ is then an arbitrary function, but any brane projection of the resulting bulk metric leads to some form of Minkowski metric. We are implicitly excluding this trivial case in all our results.}\label{trivial}
\begin{equation}\label{uderivs}
    t = u \;\;\;\;\; \psi = v\,.
\end{equation}
in the bulk metric (\ref{bulk metric}). The vector constraint (\ref{vector constraint reduced}) implies then
\begin{equation}\label{results1}
    A' =0 \;\;\;\;\;\Rightarrow \;\;\;\;\;K_{ab}=0\,,
\end{equation}
so that the scalar constraint (\ref{scalar constraint}) reduces to
\begin{equation}\label{results2}
    ^{(4)}R = \rho-3p = 0 \;\;\;\;\;\Rightarrow\;\;\;\;\;AR = constant\,.
\end{equation}

Putting all these results together, we get (after a constant factor rescaling) the following form of the bulk metric (\ref{bulk metric}) in the $k=0$ case
\begin{equation}\label{M metric}
    ds^2 = -u^{-2/3}\, (du^2 +dv^2) + u^{2/3}\,\delta_{ij}\,dx^i\,dx^j\,.
\end{equation}
Note that, in these $(u,v)$ coordinates, all metric coefficients are fully specified. This means that the general solution (\ref{M metric}) for the $k=0$ case is actually a single vacuum metric, which we will call 'M-metric' in what follows. A single 'mother' metric in the bulk for the full set of embedded spatially-flat FRW metrics, which can be recovered by projecting this M-metric onto different, infinitely-many, 4D hypersurfaces (branes).

Let us note that the M-metric (\ref{M metric}) has a time-like Killing vector
\begin{equation}\label{Killing}
    \xi \equiv \partial_v\,.
\end{equation}
This implies that the $v$-coordinate lines have an intrinsical geometrical meaning, which can be extended then to the orthogonal $u$-coordinate lines. From the physical point of view, $u \equiv R^3$ is defined by the expansion factor, meaning that the $u$-lines have an intrinsic meanig, which extend then to the orthogonal $v$-lines. This intrinsic meaning, both from the geometrical and the physical point of view, allows an straightforward comparison with other forms of the same metric. This can be even simpler if we adopt a proper-time parametrization for the $u$-lines. After some rescaling, we get
\begin{equation}\label{M metric proper}
    ds^2 = - d\tau^2 - \frac{1}{\tau}\,dv^2 + \tau\,\delta_{ij}\,dx^i\,dx^j\,.
\end{equation}
This shows explicitly how the extra time dimension collapses in the bulk as the FRW radius $R=\sqrt{\tau}$ increases during cosmic evolution.

The form (\ref{M metric proper}) is precisely the second solution obtained in the pioneering work of Ponce de Leon~\cite{JPL88} (taking $\tau = A+B\,t\,, \; v=\Psi$), and it is also the first one (exchanging the roles of $t$ and $\Psi $). A straightforward calculation shows that this is also isometric to the fourth one (taking $\tau = \Psi\,S(t),\; v=\Psi^{3/2}\,f(t)\,$)~\footnote{One recovers in this way the generic case, where the integration constant $C$ is different from zero (we get $C=1$ after a constant rescaling). The $C=0$ case, corresponding to the explicit solution given in the paper, works just for the alternative 4+1 signature}, as it should be because the M-metric is the general (non-trivial) solution: the remaining 3+2 solution in this paper corresponds actually to the trivial case that we have excluded from our analysis~\textsuperscript{1}.


\section{Trivial and non-trivial projections: Regular and Emergent Universes}

Let us consider first the trivial projection onto the $v=constant$ surfaces. This is trivial also from the geometrical point of view, as the extrinsic curvature $K_{ab}$ vanishes, so that the constraint equations (\ref{vector constraint}, \ref{scalar constraint}) amount to require $^{(4)}\!R =\rho-3p = 0$.
The resulting 4D metric can be written as
\begin{equation}\label{radiation metric}
    ds^2 = - d\tau^2 + \tau\,\delta_{ij}\,dx^i\,dx^j\,,
\end{equation}
which is the standard FRW pure radiation metric for the spatially flat case.

Of course, we have other options for projecting onto 4D hypersurfaces. We could perform for instance a linear transformation  in the $(u,v)$ plane in the bulk M-metric (\ref{M metric}) and then project onto the resulting $v'=constant$ surface. A simple calculation shows that the resulting metric would be equivalent to (\ref{radiation metric}). This amounts to say that choosing any straight line in the $(u,v)$ plane as the physical time coordinate (the one surviving in the projected brane) leads to the spatially flat FRW pure radiation metric.

Another option is to select instead a nontrivial hypersurface in order to get completely different FRW models: regular (singularity-free) universes. In the M-metric, the singularity at $u=0$ is not just a point in a time-line, but rather a line in the time plane. It is then quite easy to devise alternative time-lines which do not cross the singular line. We can select for instance hyperbolic coordinates:
\begin{equation}\label{hyperbolic coords}
    u = \psi\,e^{t}\;\;\;\;\;\; v=-\psi\,e^{-t}\,,
\end{equation}
and project onto the hypersurfaces of the form $\psi= constant > 0$. It is clear that the $u=0$ singular line is reached only asymptotically for $t\rightarrow -\infty$ (see Fig.~1): the resulting (projected) FRW metric is regular for all (finite) times.

\begin{figure}[h]
\centering
\includegraphics[width=7.8cm,height=5.0cm]{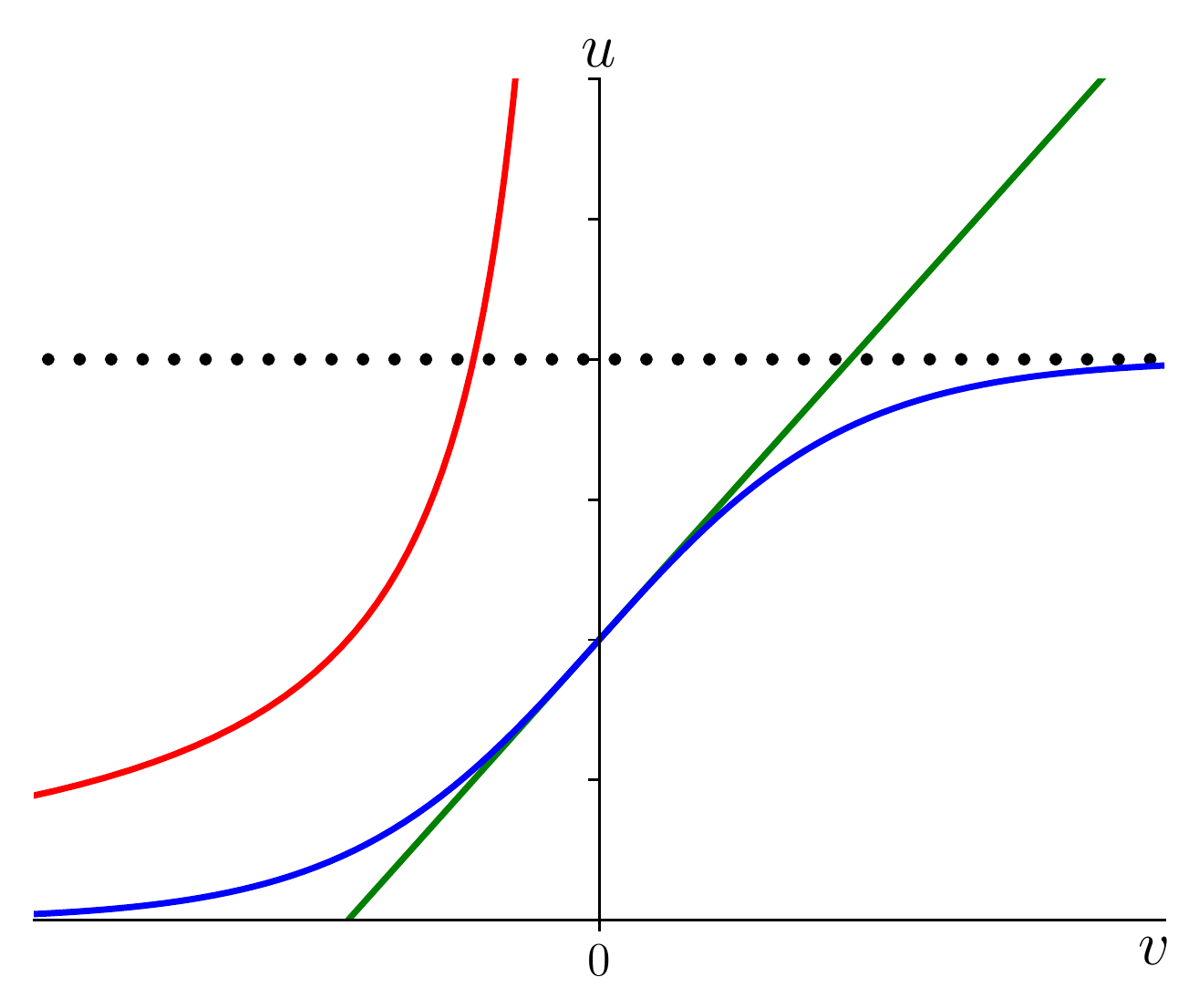}
\caption{ Timelines in the time plane of the bulk M-metric, each one leading to different FRW projected metrics. The big-bang singularity is here the $u=0$ line. The straight line corresponds to the standard pure radiation model. The other two lines correspond to FRW models without big bang, with an initial accelerating (inflationary) phase and a final decelerating phase. In the hyperbola case (red line), the deceleration is not apparent (see the text for the detailed calculation). In the hyperbolic tangent case (blue line), the inflexion point corresponds to a pure radiation phase. A stationary state is reached asymptoticaly in this case.}
\label{timelines}
\end{figure}

To be more specific, let us write down explicitly the resulting FRW model. After some constant rescaling we get:
\begin{equation}\label{regular universe}
    ds^2 = -e^{-\frac{2}{3}\,t}\,\cosh(2t)\,dt^2 + e^{\frac{2}{3}\,t}\,\delta_{ij}\,dx^i\,dx^j\,.
\end{equation}
The expansion (Hubble) factor is
\begin{equation}\label{Hubble factor}
    H = \frac{1}{3}\,e^{t/3}\,\sqrt{\sech(2\,t)}      \,,
\end{equation}
which is also finite for all times. For early times, we get an accelerated (inflationary) expansion rate, that is
\begin{equation}\label{Hubble factor early}
    H \propto R^4 \;\;\;\;\; t \ll 0\,,
\end{equation}
whereas for late times we get a deceleration in the expansion rate:
\begin{equation}\label{Hubble factor late}
    H \propto 1/R^2 \;\;\;\;\; t \gg 0\,.
\end{equation}

Another example follows from the coordinate transformation
\begin{equation}\label{hyperbolic}
    u = \psi\,[\,1+\tanh(t)\,]\;\;\;\;\;\; v=\psi\, t\,,
\end{equation}
which leads, after a constant rescaling, to the following $\psi = constant$ projection of the M-metric:
\begin{equation}\label{regular universe2}
    ds^2 = -\frac{1+\cosh(t)^{-4}}{[\,1+\tanh(t)\,]^{-2/3}}\,dt^2 + [\,1+\tanh(t)\,]^{2/3}\,\delta_{ij}\,dx^i\,dx^j\,.
\end{equation}

The resulting time curve is of course a shifted hyperbolic tangent in the time plane. We can see in Fig.~1 that there is no crossing with the $u=0$ singular line (this is again an infinite past asymptote), and that there is an upper bound for the expansion of the projected universe, which is asymptotically approached at infinite future. Note also that, as seen in Fig.~1, the physical time line is osculating to a straight line at the inflexion point ($v=0$). As the field equations contain metric derivatives just up to the second order, this matching in the metric coefficients with a radiation metric ensures that there is a radiation phase in the transition from the accelerating (inflationary) stage to the decelerating one.

In both cases, there is no beginning: the big bang singularity is just in the limit $t\rightarrow -\infty$. Although the singularity is not actually reached, the physical conditions near the singularity can be very close to those just after the big bang in standard models. This past asymptotic state can be changed by just modifying the time-line selection in the bulk. Instead of (\ref{hyperbolic coords}, \ref{hyperbolic}) we can rather choose, respectively
\begin{equation}\label{emergent1}
    u = u_0\, +\,\psi\,e^{t}\;\;\;\;\;\; v=-\psi\,e^{-t}\,,
\end{equation}
or
\begin{equation}\label{emergent2}
    u = u_0\, +\,\psi\,[\,1+\tanh(t)\,]\;\;\;\;\;\; v=\psi\, t\,,
\end{equation}
so that the curves in Fig.~1 will get displaced upwards by the amount $u_0 > 0$, safely away from the $u=0$ singularity. The corresponding FRW models will emerge then from a finite radius universe. In this way, we are building different $k=0$ approximations to the 'Emergent Universe' of Ellis and Maartens~\cite{Ellis_Maartens, Ellis}.

Of course, all these are just examples obtained by simple coordinate transformations in the bulk, leading to different brane projections. In the general case, we can define the projection by giving the time curve in explicit form, namely
\begin{equation}
    \tau = t\,,\;\; v = v(t,\Psi)\,
\end{equation}
so that the $\Psi = constant$ projection will be given by
\begin{equation}\label{projection general}
    ds^2 = - \left( 1+ \frac{\dot{v}^2}{t} \right)\,dt^2 + t\,\delta_{ij}\,dx^i\,dx^j\,,
\end{equation}
where
\begin{equation}\label{explicit}
    \dot{v} \equiv \partial_t\,v(t,\Psi)\,.
\end{equation}


\section{The completeness issue: clarifying the meaning of Campbell theorem}

Equation (\ref{projection general}) gives the brane-projected FRW metrics in explicit form, depending only on the single function (\ref{explicit}) which allows to select a specific projection of the M-metric. The corresponding expansion (Hubble) factor can then be computed explicitly:
\begin{equation}\label{Hubble general}
    H = [\,2\,t\sqrt{1+\dot{v}^2/t}\,]^{-1}\,,
\end{equation}
so that the energy density is
\begin{equation}\label{density general}
    \rho = 3H^2 = 3/4\, (t^2+t\,\dot{v}^2)^{-1}\,,
\end{equation}
which can be inverted in order to get
\begin{equation}\label{vdot}
    \dot{v}^2 = (3/4\,\rho^{-1}-t^2)/t\,.
\end{equation}
It follows from (\ref{vdot}) that there is no solution in the asymptotic limit $t \gg 0$ for the Einstein space case ($\rho = \Lambda$) nor for the pure dust case ($\rho \propto t^{-3/2}$). Moreover, it follows from (\ref{Hubble general}) that the Hubble factor cannot diverge for $t > 0$, so we cannot get 'big rip' singularities (see ref.~\cite{Jambrina sing} for a recent review).

These results go against the common belief that the Campbell theorem ensures the embedding of any four-dimensional metric into a five-dimensional Ricci-flat manifold, where the extra dimension can be either space-like or time-like. In order to clarify the real meaning of the theorem, let us point out the following:
\begin{itemize}
  \item The original theorem of Campbell~\cite{Campbell} dealt just with Riemannian manifolds. Our counter-examples are relevant only for the modern extensions to the pseudo-Riemannian case (see ref.~\cite{Seahra} for a review).
  \item The signature of the resulting 5D manifold is left unspecified, so that the 4D Ricci scalar in (\ref{scalar constraint}) appears always multiplied with a $\epsilon = \pm 1$ sign. This adds some ambiguity to the proof, as it is not clear whether the embedding works for both signs.
\end{itemize}

The key point in the proof is that the constraint equations (\ref{vector constraint}, \ref{scalar constraint}), where the 4D metric is considered as an input and the extrinsic curvature as the unknown, are under-determined. The claim is then that one should always be able to find a solution. But the under-determination in a given equation does not guarantee that there is a solution at all. In order to illustrate this, let us consider for instance a slight modification of (\ref{scalar constraint}), namely
\begin{equation}\label{scalar constraint modified}
    {K_a}^b{K_b}^a = \epsilon\;^{(4)}\!R\,
\end{equation}
which is also under-determined, but which has solution only for one of the two choices of $\epsilon$ (depending on the curvature sign), but not for the other.

Our results actually show that in the 3+2 case, where the extra dimension is time-like, the Campbell theorem holds only in its weak form: any 4D metric can be embedded in a 5D Ricci-flat manifold provided that the constraint equations hold at least in a single 4D hypersurface.


\section{Conclusions and outlook}

We have found the general solution for the embedding of (spatially flat) FRW metrics in a five-dimensional (bulk) manifold with 3+2 signature. Apart from the trivial case, the solution is unique: the M-metric (\ref{M metric}). A single 'mother' metric in the bulk for all the embedded FRW metrics.

Having a time plane in the bulk happens to be a powerful tool for devising the evolution properties of the resulting projected spacetimes: one only has to select a suitable time curve, which will be kept as the physical time coordinate in the brane. We have obtained by this method some FRW regular solutions, evolving from the infinite past (no Big Bang), that could be useful to deal with the Cosmological Horizon problem. These are regular FRW models in standard General Relativity: there is no need to recur to alternative theories in order to get these appealing cosmological models. Previous well-known cosmological solutions without big bang where obtained at the price of reducing the space symmetry group~\cite{Senovilla, Jambrina}. This is not our case: we keep the full space symmetry group; in this sense, our models pertain to the same class of the well-known 'Emergent Universe' inflationary models~\cite{Ellis_Maartens, Ellis}.

We have fully solved here only the spatially flat case (curvature index $k=0$). But note that in the general case we have reduced the full set of embedding conditions to just two equations (\ref{vector constraint reduced}, \ref{scalar constraint}) for the two functions $R(\psi,t)$ and $A(\psi,t)$. Moreover, the (infinite-dimensional) conformal group in the time plane can still be used in order to modify the expression for $A(\psi,t)$. Our conjecture is that the solution for each of the two other cases ($k=\pm 1$) is also unique, that is, that the M-metric has just one counterpart for each value of the curvature index. We will keep working in order to confirm this conjecture.

One can wonder why we have not found a flat bulk metric as an alternative starting point to the M-metric. After all, FRW metrics are known to be of embedding class one, meaning that they can actually be embedded in a flat (not just Ricci-flat) 5D manifold. Note however that in all the explicit FRW embeddings given in the classical review of Rosen~\cite{Rosen} the flat 5D manifold has four space coordinates plus only one time coordinate (4+1). Our results actually show that the lack of flat metric embeddings in the 3+2 case is not just because they are hard to find, but rather because they simply do not exist (excepting the trivial Minkowski case). This is in keeping with more recent results~\cite{Gulamov}.

Having the general solution in explicit form is crucial for solving the completeness problem: whether or not all spatially-flat FRW metrics can actually be embedded in a 3+2 bulk manifold. We have provided explicit counter-examples showing that the answer is negative. The extension of the Campbell theorem to the General Relativity case must then be considered with caution. In our case, where the extra dimension is time-like, the theorem works only in its weaker form, with the strong assumption that the constraint equations hold at least in a single 4D hypersurface.

This last result, of course, could in principle be confronted with future findings, both from the mathematical and from the physical point of view. From the mathematical point of view, any successful embedding of either the Einstein space or a pure dust spacetime (with $k=0$) in a 3+2 bulk would dismiss our counter-examples. From the physical point of view, the time-like character of the extra coordinate implies that the asymptotic future limit of the universe cannot be explained by a mixture of dust and a cosmological constant in a spatially flat geometry. It follows that either the Universe is not spatially flat ($k \neq 0$), or that the present-epoch universe (the one suggested by our current observations) must be just a transient phase. This stresses the importance of finding the general solution of the embedding problem for all values of $k$.


\acknowledgments
We acknowledge support from the Spanish Ministry of Economy, Industry and Competitiveness grants AYA2016-80289-P and AYA2017-82089-ERC (AEI/FEDER, EU). MB would like to thank CONICYT Becas Chile (Concurso Becas de Doctorado en el Extranjero) for financial support.




%
%

\bibliographystyle{prsty}

\end{document}